\def\msun{{\rm\,M_\odot}}

\def\gsim{~\rlap{$>$}{\lower 1.0ex\hbox{$\sim$}}}
\def\lsim{~\rlap{$<$}{\lower 1.0ex\hbox{$\sim$}}}
\def\etal{{\it et al.\thinspace}}
\def\wpm2{W m$^{-2}$}
\def\eg{{\it e.g.~}}

\def\ie{{\it i.e.\thinspace}}

\def\upsand{{$\upsilon$ And~}}

\documentclass[12pt, preprint]{aastex}
\usepackage{graphicx}
\usepackage{natbib,color,lscape}
\usepackage{multirow}
\usepackage{epsfig}

\title{Origin and Dynamics of the Mutually Inclined Orbits of $\upsilon$
  Andromedae c and d}
\author{Rory Barnes\altaffilmark{1,2}, Richard
  Greenberg\altaffilmark{3}, Thomas R. Quinn\altaffilmark{1}, Barbara 
E. McArthur\altaffilmark{4}, and G.~Fritz Benedict\altaffilmark{4}}

\altaffiltext{1}{Department of Astronomy, University of Washington,
Seattle, WA, 98195-1580} 
\altaffiltext{2}{Virtual Planetary Laboratory}
\altaffiltext{3}{Lunar and Planetary Laboratory, University of
Arizona, Tucson, AZ 85721}
\altaffiltext{4}{Department of Astronomy, University of Texas at Austin, TX 78712}

\begin{document}
\begin{abstract}
We evaluate the orbital evolution and several plausible origins
scenarios for the mutually inclined orbits of \upsand c and d. These
two planets have orbital elements that oscillate with large amplitudes
and lie close to the stability boundary. This configuration, and in
particular the observed mutual inclination, demands an
explanation. The planetary system may be influenced by a nearby
low-mass star, $\upsilon$ And B, which could perturb the planetary
orbits, but we find it cannot modify two coplanar orbits into the
observed mutual inclination of $30^\circ$. However, it could incite
ejections or collisions between planetary companions that subsequently
raise the mutual inclination to $>30^\circ$. Our simulated systems
with large mutual inclinations tend to be further from the stability
boundary than $\upsilon$ And, but we are able to produce similar
systems. We conclude that scattering is a plausible mechanism to
explain the observed orbits of $\upsilon$ And c and d, but we cannot
determine whether the scattering was caused by instabilities among the
planets themselves or by perturbations from $\upsilon$ And B. We also
develop a procedure to quantitatively compare numerous properties of
the observed system to our numerical models. Although we only
implement this procedure to $\upsilon$ And, it may be applied to any
exoplanetary system.
\end{abstract}

\section{Introduction}
The $\upsilon$ Andromedae ($\upsilon$ And) planetary system is the
first multiple planetary system discovered beyond our own Solar System
around a solar-like star (Butler \etal 1999). Not surprisingly, it has
received considerable attention from theoreticians, and in many ways
has been a paradigm for gravitational interactions in multiplanet
extrasolar planetary systems. At first, research focused on its
stability (\eg Laughlin \& Adams 1999; Rivera
\& Lissauer 2000; Lissauer \& Rivera 2001; Laskar 2000; Barnes \&
Quinn 2001; Go\'zdziewski \etal 2001). These investigations showed the
system appeared to lie near the edge of instability, although an
additional body could survive in between planets b and c (Rivera \&
Lissauer 2001). Later, attention turned to the apsidal motion (\eg
Stepinski \etal 2000; Chiang \& Murray 2002; Malhotra 2002; Ford \etal
2005; Barnes \& Greenberg 2006a,c, 2007a), with most investigators
considering how the apsidal behavior provides clues to formation
mechanisms such as planet-planet scattering or migration via disk
torques.  The recent direct measurement of the actual masses of
planets c and d, and especially their 30$^\circ$ mutual inclination,
via astrometry (McArthur \etal 2010) maintains this system's
prominence among known exoplanetary systems. In this investigation, we
evaluate planets c and d's gravitational interactions (ignoring b as
its orbit is still only constrained by radial velocity (RV)
observations), as presented in McArthur \etal (2010), which place strict constraints on the system's origin.

Such large mutual inclinations among planets are unknown in our Solar 
System and hence indicate different processes occurred during or after the 
planet formation process. We assume \upsand c and d formed in coplanar, 
low eccentricity orbits in a standard planetary formation models (see \eg 
Hubickyj 2010; Mayer 2010), but then additional phenomena, occurring late 
or after the formation process, altered the system's architecture. We 
consider two plausible mechanisms: a) If the distant stellar companion 
\upsand B (Lowrance \etal 2002) is on a significantly inclined (relative 
to the initial orbital plane of the planets) and/or eccentric orbit, it may 
pump up eccentricities and inclinations through ``Kozai'' interactions
(Kozai 1962; Takeda \etal 2007); or b) If the planets form close together they 
may interact and scatter into mutually inclined orbits, as shown in 
previous studies (Weidenschilling \& Marzari 2002; Chatterjee \etal 2008; 
Raymond \etal 2010).

The mutual inclinations are obviously of the most interest, but other
features of the system are also important. As \upsand represents the
first exoplanetary system with full three-dimensional orbits and true
masses directly measured (aside from the pulsar system PSR 1257+12
[Wolszczan 1994]), we may exploit this information when evaluating
formation models. To that end, we develop a simple metric that
quantifies the success of a model at reproducing numerous observed
aspects of the $\upsilon$ And planetary system. Although we apply this
approach to $\upsilon$ And, it is generalizable to any planetary
system.

In this investigation we limit the analysis to just \upsand c and d,
and to the stable fit presented in McArthur \etal (2010). As described
in that paper, the inclination and longitude of ascending node of b are not
detectable with the {\it Hubble Space Telescope}, so rather than
explore the range of architecture permitted by RV data, we focus on
the known properties of c and d. Furthermore, we do not consider the
range of uncertainties in c and d as the stable fit in McArthur et
al. (2010) is surrounded by unstable fits. As we see below, even
limiting our scope in this way, we still must perform a large number of
simulations over a wide range of parameter space.

We first ($\S$ 2) analyze the current orbital oscillations of \upsand
c and d with an N-body simulation. We then use the dynamical
properties to constrain our two inclination-raising scenarios, which
we explore through $>$50,000 N-body simulations. Then in $\S$ 3 we
show that \upsand B by itself cannot raise the mutual inclinations to
$30^\circ$. In $\S$ 4 we show that planet-planet scattering is
a likely inclination-raising mechanism. However, we find that the
most stringent constraint on scattering is the combination of its
$30^\circ$ mutual inclination \textit{and} extreme proximity
to the stability boundary.  In $\S$ 5 we discuss the results and place
them in context with previous dynamical and stability studies of
exoplanets.

\begin{center}Table 1: Current Configuration of \upsand
c and d\\
\begin{tabular}{cccccccc}
%\hline
Planet & $m$ (M$_J$) & a (AU) & e & $i$ ($^\circ$) & $\varpi$ ($^\circ$) & $\Omega$ ($^\circ$) & $n$ ($^\circ$)\\
\hline
c & 14.57 & 0.861 & 0.24 & 16.7 & 290 & 295 & 270.5\\ 
d & 10.19 & 2.70 & 0.274 &  13.5 & 240.8 & 115 & 266.1\\
\end{tabular}
\end{center}

\section{Orbital Evolution of \upsand c and d}
In this section we determine the orbital evolution of \upsand c and d.
As the mass and orbit of planet b are unknown and the four-body
interactions between \upsand A, b, c and d are extremely complex and
depend sensitively on a secular resonance, general relativity and the
stellar oblateness (McArthur \etal 2010), we have chosen to leave them
out of this analysis, but will address them in a future study. 

We examine the secular behavior of the system through an N-body
simulation using the Mercury code (Chambers 1999). Here and below we
used the ``hybrid'' integrator in Mercury. Energy was conserved to 1
part in $10^8$. For this integration we change the coordinate system
from that in the discovery paper, which is based on the viewing
geometry, and instead reference our coordinates to the invariable (or
fundamental) plane. This plane is perpendicular to the total angular
momentum vector of the system (although we ignore planetary spins),
\ie we rotated the coordinate system. The planets' orbital elements in
this coordinate system are listed in Table 1 at epoch JD
2452274.0. The system shown in Table 1 was found to be stable in
McArthur \etal, but it was also noted that the system is close to the
stability boundary. Therefore we cannot exclude the possibility that
other solutions may result in significantly different dynamical
behavior. Here and below we assume that such a situation is not the case.

\begin{figure}
\plotone{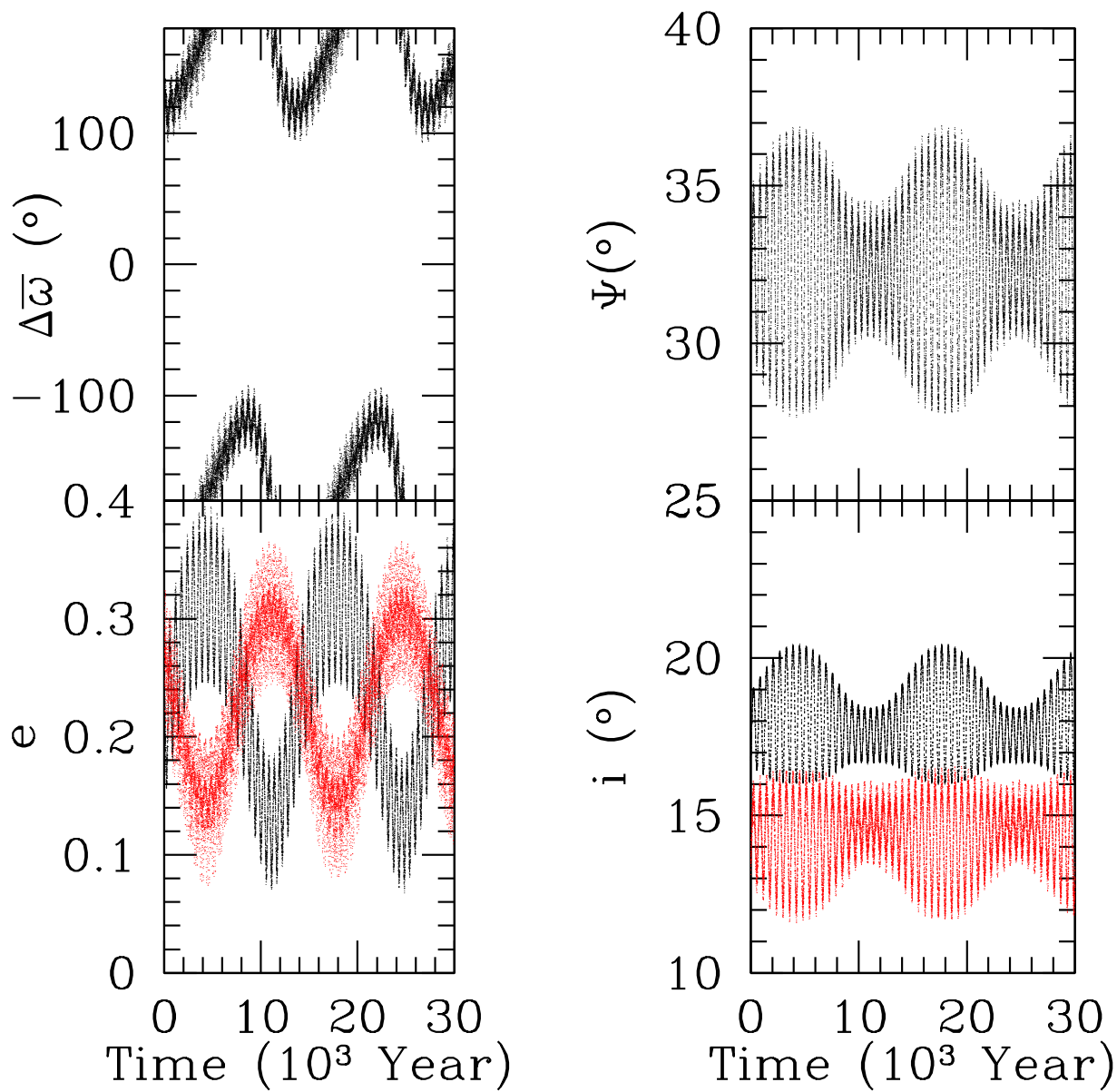}
\figcaption[]{\label{fig:secular} \small{Secular evolution of \upsand c (black)
and d (red) \textit{without} planet b. \textit{Top left:} Evolution of
$\Delta\varpi$. \textit{Bottom left:} Evolution of the
eccentricities. Black is planet c, red d. \textit{Top right:}
Evolution of $\Psi$. \textit{Bottom right:} Evolution of the
inclinations. Black is planet c, red d.}}
\end{figure}

The orbits of planets c and d undergo mutual perturbations which cause
periodic variations in orbital elements over thousands of orbits. The
long-term changes can be conveniently divided into two parts: the
apsidal evolution (changes in eccentricity $e$ and longitude of
periastron $\varpi$) and the nodal evolution (changes in inclination
$i$, longitude of ascending node $\Omega$, and hence the mutual
inclination $\Psi$). The variations, starting with the conditions in
Table 1, are shown in Fig.~\ref{fig:secular}. In this figure, we chose
a Jacobi coordinate system in order to minimize frequencies due to the
reflex motion of the star.

In Fig.~\ref{fig:secular} the left panels show the apsidal behavior,
and the right show the nodal behavior. The lines of apse oscillate
about $\Delta\varpi = \pi$, \ie anti-aligned major axes. This revision
once again changes the expected apsidal evolution. Initially Chiang \&
Murray (2002) and Malhotra (2002) found the major axes oscillated
about alignment. Then Ford \etal (2005; see also Barnes \& Greenberg
2006a,c) found the system was better described as ``near-separatrix,''
meaning the apsides lie close to the boundary between libration and
circulation. Now we find that the system found by McArthur \etal
(2010) librates in an anti-aligned sense! Substantial research has
examined the secular behavior of exoplanetary systems, yet the story
of $\upsilon$ And shows that predicting the dynamical evolution of a
planetary system based on minimum masses and poorly constrained
eccentricities is uncertain at best and foolhardy at worst. Even now,
without full three-dimensional information about planets b and e (a
trend seen in McArthur \etal [2010]) our analysis should only be
considered preliminary.

\begin{center}Table 2: Dynamical Properties of \upsand c and d\\
\begin{tabular}{cc}
%\hline
Property & Value\\
\hline
$e^{min}_c$ & 0.069\\
$e^{min}_d$ &  0.074\\
$e^{max}_c$ & 0.39\\
$e^{max}_d$ & 0.365\\
$\epsilon$ & 0.17\\
$i^{min}_c$ & $16.0^\circ$\\
$i^{min}_d$ & $11.6^\circ$\\
$i^{max}_c$ & $20.4^\circ$\\
$i^{max}_d$ & $16.7^\circ$\\
$\Psi^{min}$ & $27.6^\circ$\\
$\Psi^{max}$ & $37.1^\circ$\\
$\beta/\beta_{crit}$ & 1.075\\ 
\end{tabular}\\
\end{center}

The eccentricities and inclinations undergo large oscillations, as
does the mutual inclination. The slow $\sim$15,000 year oscillations
are expected from analytical secular theory. The high-frequency
oscillation is
probably a combination of coupling between eccentricity
and inclination and velocity changes that occur during conjunction
(note that the impulses at each conjunction can change $e$ by more
than 0.01). Note also that the ratio of the orbital period, 5.32,
is not very close to the low-order 5:1 and 11:2 resonances, so
this high frequency evolution is not due to a mean motion resonance.

The numerical integration also allows a calculation of the proximity
to the apsidal separatrix $\epsilon$ (Barnes \& Greenberg 2006c). When
$\epsilon$ is small ($\lsim 0.01$), two planets are near the boundary
between librating and circulating major axes. Although $\upsilon$ And
was the first system to be identified as near-separatrix (Ford \etal
2005), we find $\epsilon = 0.17$ indicating that the system is
actually {\it not} close to the separatrix, according to the McArthur \etal
(2010) model.

Also of interest is the system's proximity to dynamical
instability, \eg the ejection of a planet. Previous studies found
planets c and d are close to this limit (Rivera \& Lissauer 2000;
Barnes \& Greenberg 2001, 2004; Go\'zdziewski \etal 2001). Here we
calculate c and d's proximity to the Hill stability boundary with the quantity
$\beta/\beta_{crit}$ (Barnes \& Greenberg 2006b,2007b; see also
Marchal \& Bozis 1982; Gladman 1993; Veras \& Armitage 2004). If
$\beta/\beta_{crit} > 1$ then the pair is stable, if $<1$, it is
unstable. We find $\beta/\beta_{crit} = 1.075$ and therefore these two
planets are very close to the dynamical stability boundary. We caution
that constraints based on $\beta/\beta_{crit}$ may be misleading, as
Hill stability is strictly only applicable to three-body systems.

In Table 2 we list some statistics of our $10^6$ year integration
based on 36.5 day output intervals in astrocentric
coordinates. Superscripts $min$ and $max$ refer to the minimum and
maximum values achieved, respectively. Clearly the actual dynamics of
this system depend on the presence and properties of planet b, and, in
principle, any additional unconfirmed planets, but without better data
on these objects, Table 2 is the best available characterization of
the orbital evolution of these two planets. They may also be used to
evaluate the origins scenarios described in the following two
sections.

\section{Perturbations from \upsand B}
\upsand B is a distant M4.5, 0.2 $\msun$ companion star 
to \upsand A (Lowrance \etal 2002; McArthur \etal 2010). Its
orbit can not be estimated yet, hence we do not know if it is
gravitationally bound (Patience \etal [2002]; Raghavan \etal
[2006]). Estimates for their separation range from 702 AU (Raghavan
\etal 2006) to as much as 30,000 AU (McArthur \etal 2010). If planet c
and d formed on circular, coplanar orbits, then could
$\upsilon$ And B have pumped up the mutual inclinations of c and d?

Given the uncertainty in B's orbit, we consider a broad parameter
space sweep: 13,200 simulations that cover the range $500 \le a_B \le
2000$ AU, $0.5 \le e_B \le 0.85$ and $30^\circ \le i_B \le
80^\circ$. Here $i_B$ is referenced to the initial orbital plane of
the planets, not the invariable plane of the four-body system. The
angular elements were varied uniformly from 0 to 2$\pi$. Note that
this range was chosen to increase the perturbative effects of B and
does not reflect any expectation of its actual orbit. Each simulation
was run for $5 \times 10^6$ years, which corresponds to $\sim 250$
orbits of B. While not a long time, we find that B may destabilize the
planetary system, which could lead to large mutual inclinations of
planet c and d (see $\S$ 4).

For the vast majority of these cases, the orbits of c and d remain
coplanar, with $\Psi < 1^\circ$. However, 3 simulations ejected planet
d; 192 led to planets with $\Psi^{max} > 1^\circ$; 26 with $\Psi^{max}
> 10^\circ$; and 1 case out of the 13,200 reached $\Psi^{max} =
34^\circ$. The 192 non-planar cases were spread throughout parameter
space, with no significant clustering.

To explore effects on longer timescales, we integrated 20 cases to 1
Gyr. Four of the previous simulations that had led to significant
mutual inclinations (including that which led to $\Psi^{max} =
34^\circ$) were tested, in order to examine stability. The other 16
were chosen from among those in which $\Psi^{max}$ stayed $<1^\circ$
over 250 orbits of B, in order to determine if mutual inclinations
could develop over longer timescales. We find that most of these
simulations, in fact, ejected a planet. Therefore the orbit of \upsand
B appears able to destabilize a circular, coplanar system.

\begin{figure}
\plotone{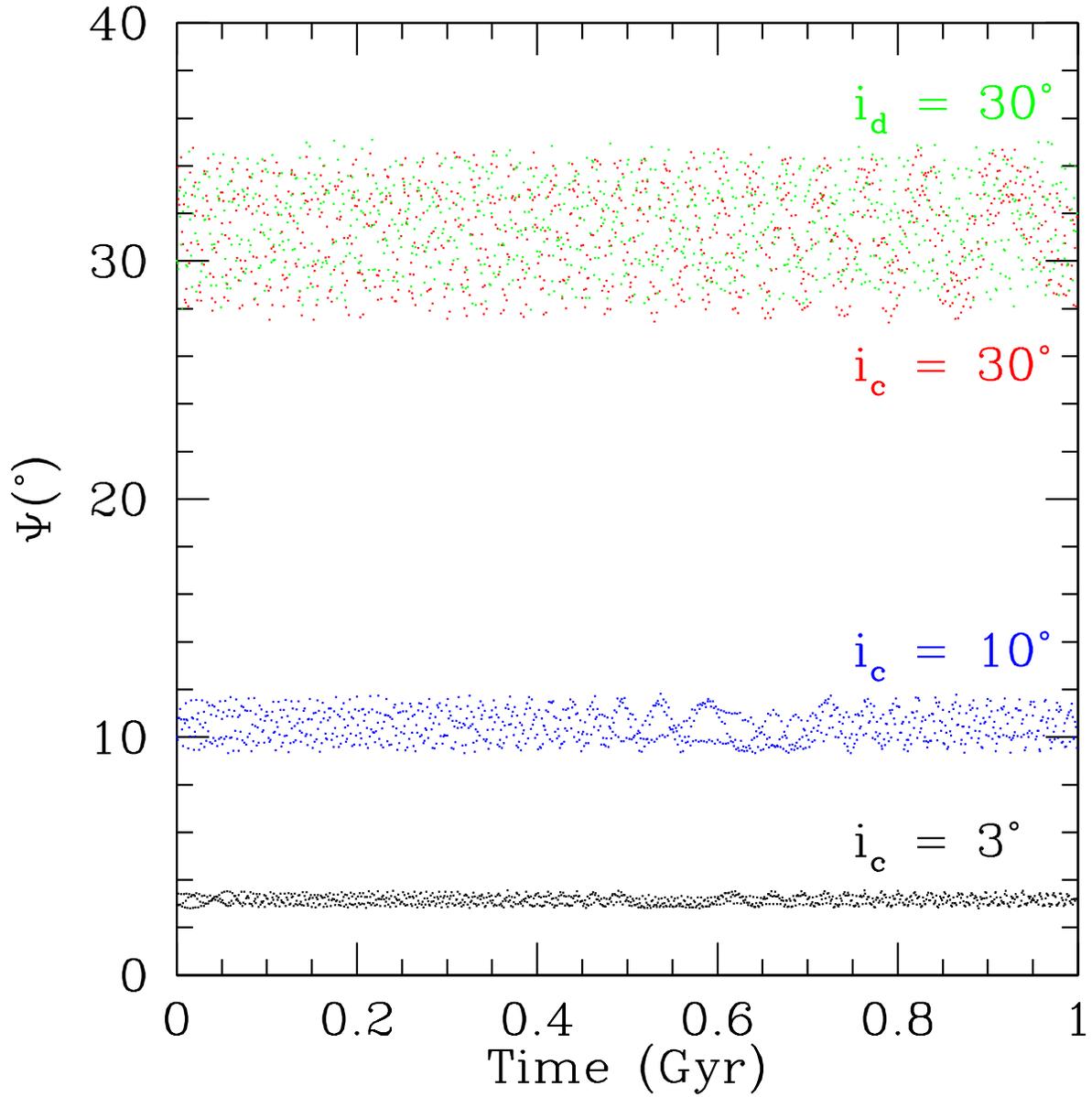}
\figcaption[]{\label{fig:kozainonzero} \small{Variation of $\Psi$ for four
hypothetical systems with \upsand B included (see text for more
details). The black points are systems in which initially $i_c = 3^\circ$, blue $i_c = 10^\circ$, red $i_c = 30^\circ$, and green $i_d = 30^\circ$.}}
\end{figure}

Next we relax the requirement that the planets began on coplanar
orbits and ran simulations with initial mutual inclinations $\Psi_0 =
3^\circ$, $10^\circ$, and $30^\circ$, and with $a_B = 700$, $e_B = 0$,
$i_B = 30^\circ$. The two planets began with their current best fit
semi-major axes and masses, but on circular orbits, and one
inclination was set to $3^\circ$, $10^\circ$ or $30^\circ$ with the
masses held constant. These systems were integrated for 1 Gyr.  In
each of these cases, shown in Fig.~\ref{fig:kozainonzero}, the initial
value of $\Psi$ is maintained for the duration of the simulation.  The
widths of the libration increase with $\Psi_0$ because the
interactions among the planets are driving a secular oscillation. It
therefore seems that even if the planets began with a nonzero relative
inclination, \upsand B is unable to pump it to the range shown in
Fig.~\ref{fig:secular}. These simulations also demonstrate that
plausible orbits of \upsand B will not destabilize the observed
system.

These simulations indicate that it is unlikely that \upsand B could
have twisted the orbits into the mutually inclined system we see
today. However, it could have destabilized the system, which we will
see in the next section is a process which can lift coplanar orbits to
$\Psi \gsim 30^\circ$.

\section{Planet-Planet Scattering}
Our second hypothesis considers the possibility that the planetary
system formed in an unstable configuration independent of B, and that
encounters between planets ultimately ejected an original companion
leaving a system with high mutual inclinations. Such impulsive
interactions could drive inclinations to large values, perhaps as
large as $60^\circ$ (Marzari \& Weidenschilling 2002; Chatterjee \etal
2008; Raymond \etal 2010). We considered 41,000 different initial
configuration of the system, \eg one or two additional planets with
masses in the range 1 -- 15 M$_{\textrm{Jup}}$. At the end of this
section we summarize the results of all these simulations, but
initially we focus on one subset.

We completed 5,000 simulations that began with three 10 -- 15
M$_{\textrm{Jup}}$ mass objects (uniformly distributed in mass)
separated by 4--5 mutual Hill radii (Chambers \etal 1996), with $e <
0.05$, $i < 1^\circ$ and $0.75 < a < 4$ AU. We integrated these cases
for $10^6$ years with Mercury's hybrid integrator, conserving energy
to 1 part in $10^4$. About 1\% of cases failed to conserve energy at
this level and were thrown out. These ranges are somewhat arbitrary
but follow the recent study by Raymond \etal (2010), which considered
smaller mass planets at larger distances. They found that a system
consisting of three 3 M$_{\textrm{Jup}}$ planets could, after removal
of one planet and settling into a stable configuration, end up with
$\Psi > 30^\circ$ about 15\% of the time (down from 30\% for three
Neptune-mass planets). However, they also found that only 5\% of
systems of three 3 M$_{\textrm{Jup}}$ planets settled into a
configuration with $\beta/\beta_{crit} < 1.1$. Therefore we expect
that these two parameters will be the hardest to reproduce via
scattering. As we see below, this expectation is borne out by our
modeling.

In our study, a successful model conserved energy adequately (1 part
in $10^4$), removed the extra planet, and the remaining planets all
had orbits with $a < 10$ AU.  2072 trials met these requirements (1416
collisions and 656 ejections). We ran each of these final two-planet
configurations for an additional $10^5$ years (again validating the
simulation via energy conservation) to assess secular behavior for
comparison with the system presented in Table 2.

\begin{figure}
\plotone{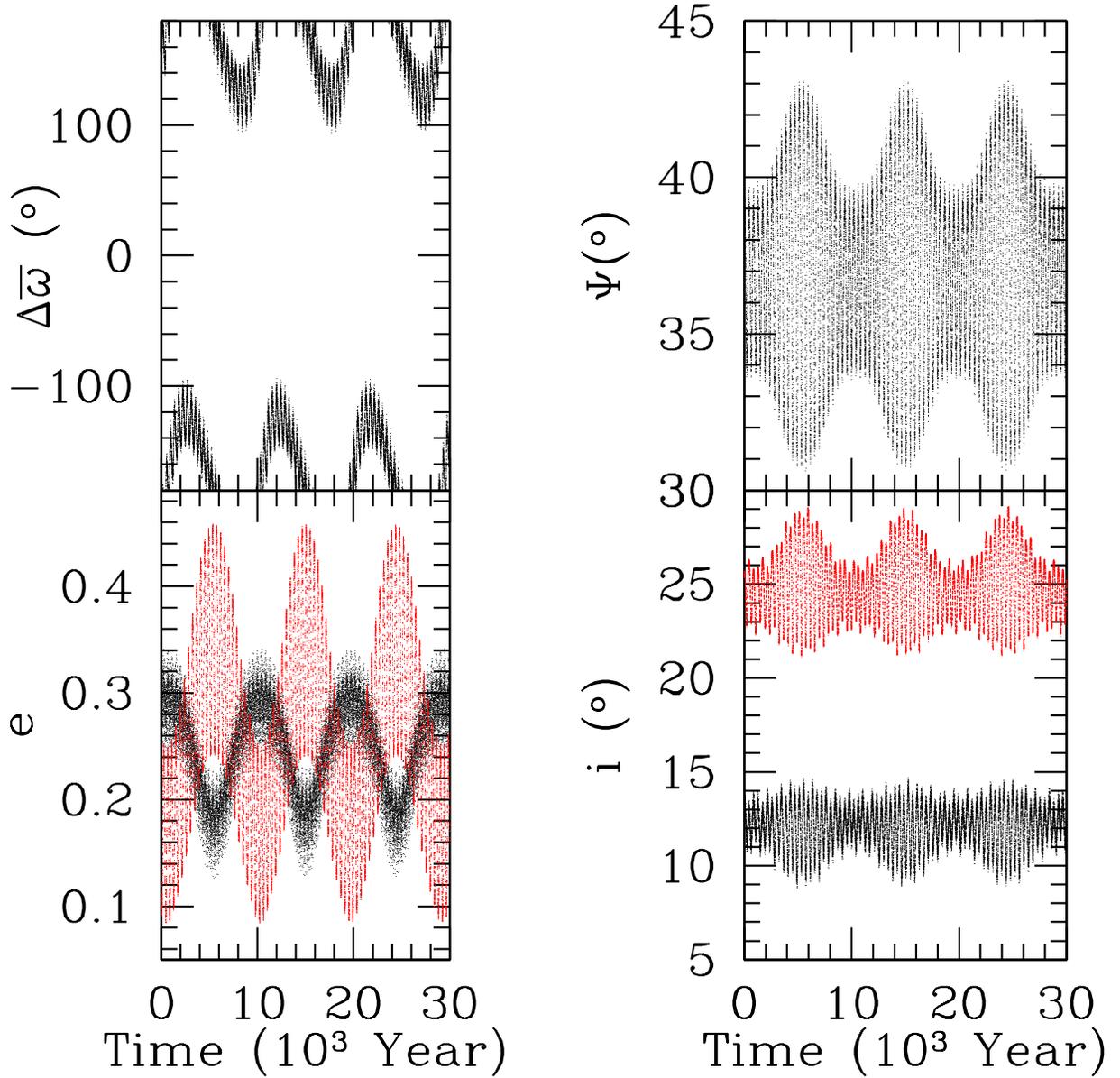}
\figcaption[]{\label{fig:example} \small{Secular evolution of a
simulated system after the ejection of an exterior planet.\textit{Top
left:} Evolution of $\Delta\varpi$. \textit{Bottom left:} Evolution of
the eccentricities. Black is planet c, red d. \textit{Top right:}
Evolution of $\Psi$. \textit{Bottom right:} Evolution of the
inclinations.}}
\end{figure}

In Fig.~\ref{fig:example} we show the outcome of one such trial in which a 
hypothetical planet was ejected. The format of this figure is the same as 
Fig.~\ref{fig:secular}. The behavior is qualitatively similar as in 
Fig.~\ref{fig:secular}, including anti-aligned libration of the apses, the 
magnitudes of the eccentricities and the inclinations, and the short 
period oscillation superposed on the longer oscillation. The mutual 
inclination for this case is even larger than the observed system. This 
system's $\beta/\beta_{crit}$ is 1.06, slightly lower than the observed 
system. This simulation, which is one of the closest matches to the 
observed system, shows that the ejection of a single additional 
planet could have produced the \upsand system.

\begin{figure}
\epsscale{0.85}
\plotone{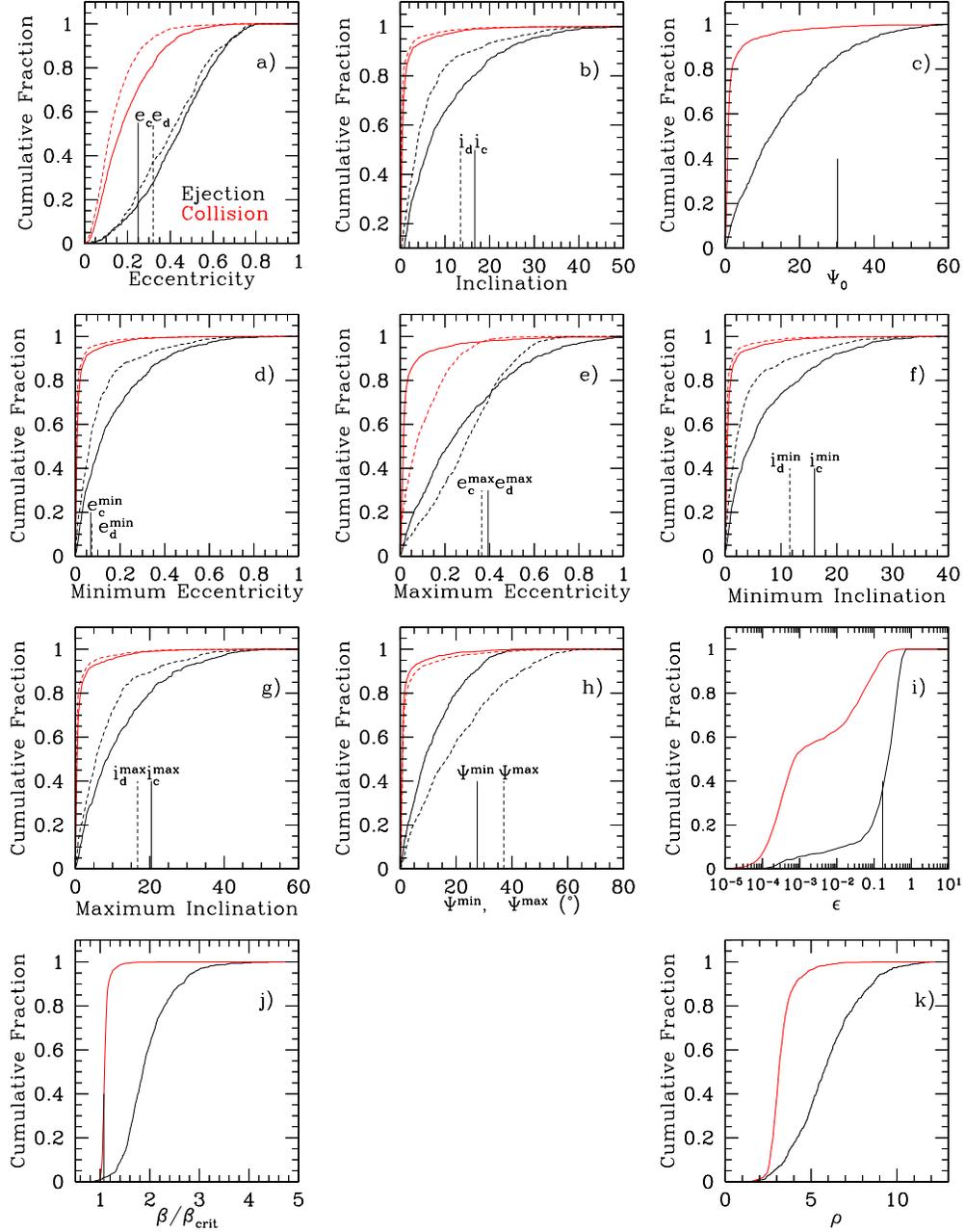}
\caption{\label{fig:single}Cumulative distributions of properties (out
of 626 ejections, 1416 collisions) after 
scattering of one planet. Black curves refer to cases in which one 
planet was ejected, red to a collision between two planets. Thick vertical 
lines correspond to the value for the best fit system. If the parameter is 
measurable for both planets, solid lines indicate planet c, dashed d.}
\end{figure}

Figure \ref{fig:example} is but one outcome. We next explore the 
statistics of this suite of simulations and consider the other orbital 
elements and dynamical properties. We divide the outcomes into two cases: 
Ejections and Collisions. These two phenomena could produce significantly 
different outcomes. For example, collisions tend to occur near periastron 
of one planet and apoastron of the other, and we might expect the merged 
body to have a lower eccentricity than either of the progenitors. We show 
the cumulative distributions of the properties listed in Table 2 in 
Fig.~\ref{fig:single}. Comparing the values of orbital elements at a given 
time is not ideal, but as it has been done many times (see \eg Ford \etal 
2001; Ford \& Rasio 2008; Juric \& Tremaine 2008; Chatterjee \etal 2008; 
Raymond \etal 2010), we do so here as well. In Figs.~\ref{fig:single}a--c, 
we show the values of $e$, $i$, and $\Psi$ at the end of the initial 
$10^5$ year integration.

In panels d--h we show the ranges of $i^{min}$, $i^{max}$,
$\Psi^{min}$ and $\Psi^{max}$. We find that 8.9\% of successful models
produced a system with $\Psi^{max} > 30^\circ$, consistent with
Raymond \etal (2010). Note that ejections produce $\Psi^{max} >
30^\circ$ about 20\% of the time.

In Fig.~\ref{fig:single}i we show the $\epsilon$ distribution, which
is bimodal with one peak near 0.1 and another near $10^{-3}$. The
observed value of 0.17 is not an unusual value, and we find that
systems with this $\epsilon$ value can have appropriate values of
$e^{min}$ and $e^{max}$. We note that the significant fraction of
systems near the apsidal separatrix contradicts the results of Barnes
\& Greenberg (2007a), which found that scattering only produced
$\epsilon < 0.01$ a few percent of the time. The most likely
explanation for this difference is that Barnes \& Greenberg considered
coplanar orbits and forbade collisions, whereas here we explore
non-planar motion.  Our results also indicate that near-separatrix
motion is likely a result of collisions, rather than ejections, and
$\epsilon < 10^{-4}$ (which is unlikely to be measured any time soon)
only result from collisions.

Figure \ref{fig:single}j shows the distribution of
$\beta/\beta_{crit}$ after scattering. Here the difference between
collisions and ejections is starkest: Ejections have a much broader
distribution than collisions. Barnes \etal (2008) noted that systems
are ``packed'' (no additional planets can lie in between two planets)
when $\beta/\beta_{crit} \lsim 2$, which is near the peak of the
ejection distribution. Our results consistent with Raymond \etal (2009).

In this model the mutual inclinations and proximity to instability are the
strongest constraints on the system's origins, but we may quantify
scattering's ability to reproduce all the observed features of the
system. Most previous studies of scattering have focused on
reproducing eccentricity distributions, but all available information
should be used. With this goal in mind, we lay out here a simple
method to quantify any model's ability to reproduce observations. For
each simulated system, we calculate its ``parameter space
distance'' $\rho$ from the best fit. We define this quantity as
\begin{equation}
\label{eq:rho}
\rho = \sqrt{\sum{\Big(\frac{\eta - \eta_j}{\eta}\Big)^2}},
\end{equation}
where $\eta$ represents $e^{min}_j...\beta/\beta_{crit}$ and $j$ =
c,d (see Table 2). This statistic has several limitations: It ignores
correlations between parameters; is dependent on the coordinate
system; ignores uncertainties in the observations (as discussed
above), and possibly overweights some parameters by including
combinations of variables that are not independent. Although crude,
$\rho$ does provide a quantitative estimate of how close a modeled
system is to the observed system. Smaller values of $\rho$ signal a
system that is a closer match to the actual system.

In Fig.~\ref{fig:single}k we plot the distributions of $\rho$. These
distributions resemble the $\beta/\beta_{crit}$ distributions (panel
j), suggesting it is the most important constraint on the system.  In
Table 3 we show some statistics of our runs, where ``min'' is the
minimum value of the set, ``avg'' is the mean, $\sigma$ is the
standard deviation, and ``max'' is the maximum value.

\begin{figure}
\plotone{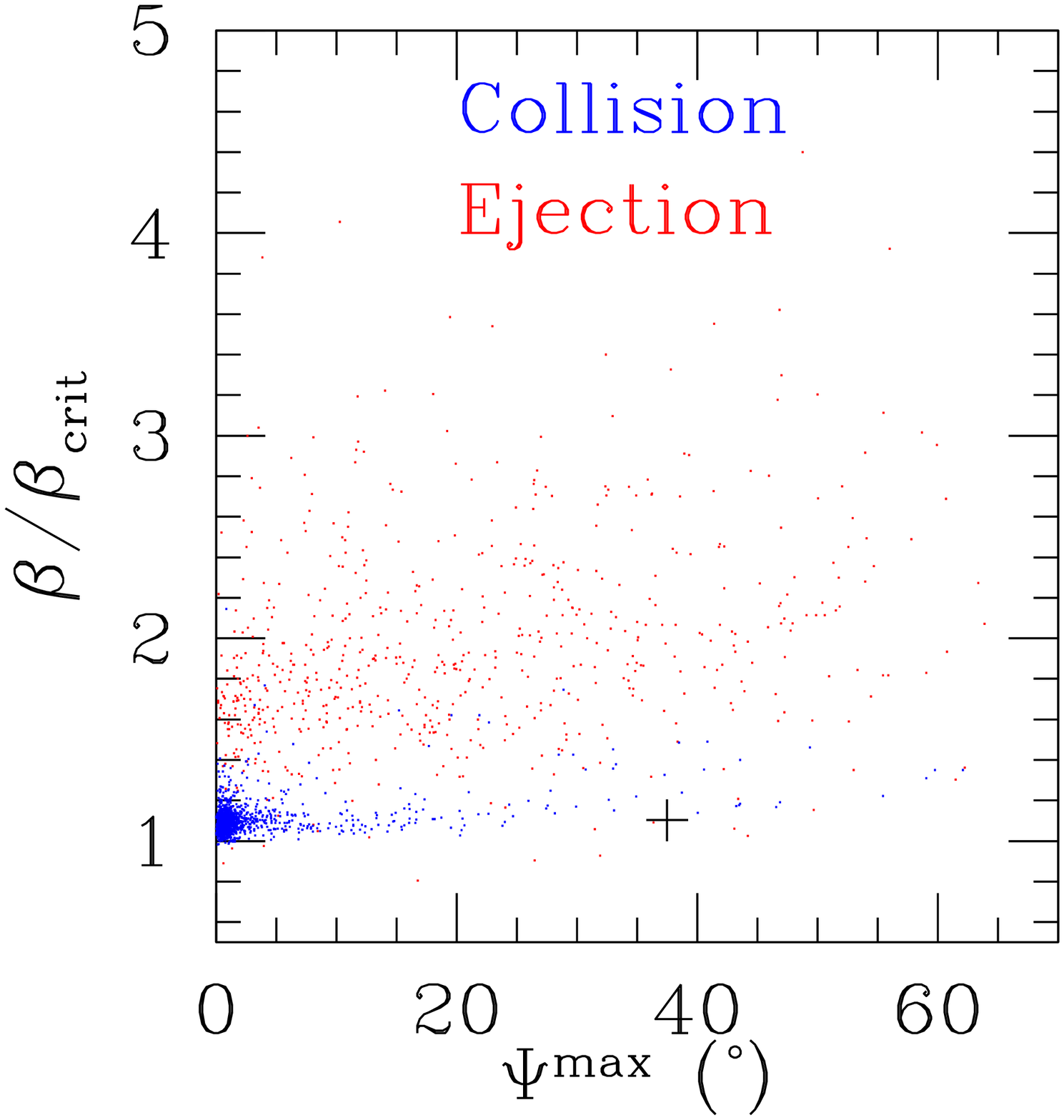}
\figcaption[]{\label{fig:psibbc} \small{Relationship between proximity
to instability ($\beta/\beta_{crit}$) and maximum mutual inclination
($\Psi^{max}$). Systems which experienced a collision are color-coded
blue, ejections red. The best fit to the $\upsilon$ And system is
shown by the + sign.}}
\end{figure}

For most of the parameters we consider, ejections and collisions do a
reasonable job of producing the observed values. However, this
representation does not show any cross-correlation, \ie does a system
with high $\Psi$ also have low $\beta/\beta_{crit}$? We explore this
relationship in Fig.~\ref{fig:psibbc}. We see that post-collision
systems (blue points) cluster heavily at low $\beta/\beta_{crit}$ and
$\Psi^{max}$, but post-ejection systems (red points) have a much
broader range. Nonetheless, the two outcomes seem
equally likely to reproduce the system, represented by the ``+''
(recall that there are three times as many blue points as
red). However, from our models the actual probability that
instabilities can reproduce the $\upsilon$ And system is less than
1\%.

From our analysis of these 2072 systems, we see that it is possible
for planetary ejections and collisions to reproduce the observed
$\upsilon$ And system, albeit with low probability. This suite of
simulations is obviously limited in scope, so we performed 36,000 more
simulations relaxing constraints on planetary mass (allowing uniform
values between 1 and 15 M$_{\textrm{Jup}}$), separation (uniform
distribution between 2 and 5 mutual Hill radii), and number of planets
(3 or 4). These other simulations began with two planets with
approximately the same semi-major axes as observed, and placed planets
interior and/or exterior to these two planets. These simulations show
that the equal-mass case we considered here is the best method to
achieve large $\Psi$, as expected from Raymond \etal (2010). For
additional planets with masses less than 5 M$_{\textrm{Jup}}$, $\Psi$
values greater than $30^\circ$ are very unlikely, but
$\beta/\beta_{crit} \sim 1$ is more likely. The removal of two planets
does not make much difference in the resulting system. We conclude
that the removal of 1 or more planets with mass(es) larger the 5
M$_{\textrm{Jup}}$ is a viable process to produce the observed
configuration of $\upsilon$ And c and d.

\begin{center}Table 3: Distribution of Properties After Collision/Ejection
\begin{tabular}{c|cccc|cccc}
%\hline\hline
Property & \multicolumn{4}{|c|}{Ejection} & \multicolumn{4}{|c}{Collision}\\
\hline
 & min & avg & $\sigma$ & max & min & avg & $\sigma$ & max\\
$e^{min}_c$ & $3.5 \times 10^{-5}$ & 0.29 & 0.17 & 0.74 & $6 \times 10^{-6}$ & 0.1 & 0.1 & 0.64\\
$e^{min}_d$ & $7.6 \times 10^{-5}$ & 0.28 & 0.19 & 0.75 & $8 \times 10^{-6}$ & 0.03 & 0.07 & 0.51\\
$e^{max}_c$ & 0.097 & 0.53 & 0.16 & 0.83 & 0.024 & 0.26 & 0.15 & 0.79\\
$e^{max}_d$ & 0.064 & 0.28 & 0.16 & 0.88 & 0.05 & 0.25 & 0.12 & 0.77\\
$\epsilon$ & $10^{-4}$ & 0.26 & 0.22 & 1.36 & $3.7 \times 10^{-5}$ & 0.066 & 0.12 & 1.1\\
$i^{min}_c$ ($^\circ$) & 0.02 & 7.2 & 7.50 & 34.4 & 0.013 & 1.2 & 2.7 & 30.4\\
$i^{min}_d$ ($^\circ$) & 0.002 & 4.1 & 5.9 & 43.7 & 0.005 & 0.77 & 2.1 & 30.4\\
$i^{max}_c$ ($^\circ$) & 0.08 & 11.9 & 10.8 & 50.8 & 0.014 & 1.7 & 3.8 & 43.0\\
$i^{max}_d$ ($^\circ$) & 0.03 & 8.3 & 8.7 & 48.7 & 0.007 & 1.26 & 3.4 & 49.7\\
$\Psi^{min}$ ($^\circ$) & 0.06 & 11.7 & 9.7 & 46.8 & 0.036 & 2.0 & 4.5 & 39.8\\
$\Psi^{max}$ ($^\circ$) & 0.1 & 19.8 & 15.5 & 63.9 & 0.032 & 2.9 & 6.7 & 62.1\\
$\beta/\beta_{crit}$ & 0.8 & 1.95 & 0.49 & 4.4 & 0.98 & 1.11 & 0.08 & 2.14\\
$\rho$ & 1.17 & 5.9 & 1.9 & 11.9 & 1.08 & 3.3 & 0.75 & 8.6\\
\end{tabular}
\end{center}

\section{Conclusions}
We have shown that the orbital behavior of the model for Ups And c and
d proposed by McArthur et al. (2010) is quite different from that of
orbital models identified by previous studies that had no knowledge of
the inclination or actual mass of the planets. The major axes librate
in an anti-aligned configuration, and their mutual inclination is
substantial and oscillates with an amplitude of about $10^\circ$.

We find that the companion star \upsand B by itself cannot pump the
mutual inclination up to large values, even if the planets began with
a significant relative inclination. However, it may have sculpted the
planetary system by inciting an instability that ultimately led to
ejections of formerly bound planets. The timescale to develop these
instabilities is long. The configurations of B that could have such
effects are sparsely distributed over parameter space, and the orbits
of previously bound planets cannot be specified. These factors make
the role of \upsand B complicated, but suggest an in-depth analysis of
its role merits further research.

Even without B, planet-planet scattering may have driven the system to
the observed state. That process can easily reproduce the apsidal
motion, but pumping the mutual inclination up to the observed values
is difficult and probably requires the removal of a planet with mass $> 5$ 
M$_\textrm{Jup}$. Removing two planets does not increase
this probability significantly.

The other important constraint on the scattering hypothesis is the
system's close proximity to the stability boundary,
$\beta/\beta_{crit}$. Collisions may leave a system near that
boundary, whereas ejections tend to spread out the
planets. Furthermore, we find that collisions tend to produce systems
with low $\beta/\beta_{crit}$ and low $\Psi$, while ejections produce a
broad range of $\Psi$, but large values of
$\beta/\beta_{crit}$. Nonetheless, Fig.~\ref{fig:example} demonstrates
that scattering can produce systems similar to $\upsilon$ And.

Although scattering is a reasonable process to produce the observed
architecture, we cannot determine the triggering mechanism. Did
scattering occur because \upsand B destabilized the planetary system?
Or did the planet formation process itself, independent of B,
ultimately lead to instabilities? The presence of \upsand B makes
distinguishing these possibilities very difficult. A larger census of
mutual inclinations and stellar companions can resolve this open
issue.

Alternatively, our decisions about the system at the onset of
scattering could be mistaken. We assumed the planets formed inside the
original protoplanetary disk with inclinations $<1^\circ$. It may be
that larger initial inclinations are possible prior to scattering, in
which case the planets could be pumped to larger mutual inclinations
(Chatterjee \etal 2008). However, it remains to be seen if such
configurations are possible prior to scattering. Future studies should
explore the inclinations of giant planets during formation.

We have also ignored the effects of planet b, stellar companion B, and a 
possible fourth planetary companion (McArthur \etal 2010) in our analysis. 
These bodies could significantly change the secular behavior, and/or the 
observed fundamental plane. Furthermore, planet b is tidally interacting 
with its host star, which could alter the long-term secular behavior (Wu 
\& Goldreich 2002). Hence future revisions to this system, and the 
inclusion of tidal effects, could significantly alter the
interpretations described above, possibly making scattering more
likely to produce the observed system. We are currently exploring the
wide range of $i$'s and $\Omega$'s of b and the subsequent orbital
evolution of the entire system.

Figure \ref{fig:single}i shows that scattering tends to produce two
types of apsidal behavior: near-separatrix ($\epsilon < 10^{-3}$) and
motion far from the separatrix ($\epsilon > 0.01$) with a desert in
between. Adding a second scatterer to the mix does not erase this
bimodality. This result contrasts with the case with no inclinations
(Barnes \& Greenberg 2007a), in which near-separatrix motion
($\epsilon < 0.01$) is not a common outcome of
scattering. Fig.~\ref{fig:single}i shows that, in fact, both outcomes
are likely, at least in systems similar to $\upsilon$ And. Although
there are hints of this structure in the observed exoplanet population
(Barnes \& Greenberg 2006c), those results are based on radial
velocity data. We now know that minimum masses are not necessarily a
good indicator of apsidal motion.

For $\upsilon$ And, the large mutual inclination and proximity to
instability are strong constraints on the origin of its planetary
system. However, for other systems, this may not be the case. We
outlined in $\S$ 4 a method in which all aspects of a planetary system
can be combined to quantify the validity of a formation model. When
the mass and three dimensional orbits of a planetary system are known,
the properties presented in Table 1 can be combined into a single
parameter $\rho$ which provides a statistic for quantitatively
comparing models. In our analysis we ignored the observational errors,
which is regrettable, but necessary due to the system's extreme
proximity to dynamical instability (McArthur \etal 2010). We encourage future
studies that strive to reproduce the system of McArthur \etal to find $\rho$ values less that those listed in Table 3, as
lower values imply a closer match to the system, assuming that other stable
solutions show similar behavior to the one we describe here.  Furthermore,
investigations into exoplanet formation could compare distributions of
observed and simulated properties as a quantitative method for model
validation.

The revisions of McArthur \etal (2010) reveal the importance of the
mass-inclination degeneracy in dynamical studies of
exoplanets. Clearly in some cases masses can be much larger than the
minimum value measured by radial velocity, which in turn changes
secular frequencies and eccentricity amplitudes. However, large
changes in mass due to the mass-inclination degeneracy should be
rare, hence, trends using minimum masses may still be valid. Nonetheless, we urge
caution when exploring trends among dynamical properties (\eg
Zhou \& Sun 2003; Barnes \& Greenberg 2006c), as they may be
misleading.

Even if $30^\circ$ mutual inclinations turn out to be rare, systems
with $\Psi \sim 10^\circ$ probably are not (Fig.~\ref{fig:single}; Chatterjee \etal 2008; Raymond \etal 2010). If these
systems host planets with habitable climates, they may be very
different worlds than Earth. Planetary inclinations can drive
obliquity variations in terrestrial planets (Atobe \etal 2004,
Armstrong \etal 2004; Atobe \& Ida 2007), unless they have a large
moon (Laskar 1997). Therefore future analyses of potentially habitable
worlds should pay particular attention to the mutual inclinations, and
climate modeling should explore the range of possibilities permitted
by large mutual inclinations. Terrestrial planets will likely be
discovered in their star's habitable zone in the coming years. The
orbital configuration and evolution of $\upsilon$ And warns us that
habitability assessment hinge on the orbital architecture of the
entire planetary system.

\medskip

RB acknowledges support from the NASA Astrobiology Institute's Virtual 
Planetary Laboratory lead team, supported by cooperative agreement No. 
NNH05ZDA001C. RG acknowledges support from NASA's Planetary Geology 
and Geophysics program, grant No. NNG05GH65G. BEM and GFB 
acknowledge support from NASA through grants GO-09971, GO-10103, and 
GO-11210 from the Space Telescope Science Institute, which is operated by 
the Association of Universities for Research in Astronomy (AURA), Inc., 
under NASA contract NAS5-26555. We also thank Sean Raymond for helpful discussions.

\references
Armstrong, J.C., Leovy, C.B., \& Quinn, T.R. 2004, Icarus, 171, 255\\
Atobe, K., Ida, S. \& Ito, T. 2004, Icarus, 168, 223\\
Atobe, K., Ida, S. 2007, Icarus, 188, 1\\
Barnes, R., Go\'zdziewski, K., \& Raymond, S.N. 2008, ApJ, 680, L57\\
Barnes, R. \& Greenberg, R. 2006a, ApJ, 652, L53\\
------------. 2006b, ApJ, 638, 478\\
------------. 2006c, ApJ, 647, L163\\
------------. 2007a, ApJ, 665, L67\\
------------. 2007b, ApJ, 659, L53\\ 
Barnes, R. \& Quinn, T.R. 2001, ApJ, 554, 884\\
------------. 2004, ApJ, 611, 494\\
Barnes, R. \& Raymond, S.N. 2004 ApJ, 617, 569\\
Butler, R.P. \etal 1999, ApJ, 526, 916\\
Chambers, J., 1999, MNRAS, 304, 793\\
Chatterjee, S. \etal 2008, ApJ, 686, 580\\
Chiang, E. I., Tabachnik, S., \& Tremaine, S. 2001, AJ, 122, 1607\\
Ford, E.B., Lystad, V., \& Rasio, F.A. 2005, Nature, 434, 873\\
Ford, E.B. \& Rasio, F.A. 2008, ApJ, 686, 621\\
Gladman, B. 1993, Icarus, 106, 247\\
Go\'zdziewski, K. \etal 2001, A\&A, 378, 569\\
Hubickyj, O. 2010, in \textit{Formation and Evolution of Exoplanets}, Rory Barnes (ed), Wiley-VCH, Berlin.\\
Juri\'c, M. \& Tremaine, S. 2008, ApJ, 686, 603\\
Kozai, Y. 1962, AJ, 67, 591\\
Laskar, J. 2000, PhRvL, 84, 3240\\
Laskar, J., Joutel, F., \& Robutel, P. 1993, Nature, 615\\
Laughlin, G. \& Adams, F.C. 1999, ApJ, 526, 881\\
Lissauer, J.J. \& Rivera, E.J. 2001, ApJ, 554, 1141\\
Lowrance, P.J., Kirkpatrick, J.D., \& Beichman, C.A. 2002, ApJ, 572, L79\\
Malhotra, R. 2002, ApJ, 575, L33\\
Marchal, C. \& Bozis, G. 1982, CeMDA, 26, 311\\
Marzari, F. \& Weidenschilling, S. 2002, Icarus, 156, 570\\
Mayer, L. 2010 in \textit{Formation and Evolution of Exoplanets}, Rory Barnes (ed), Wiley-VCH, Berlin.\\
McArthur, B. \etal 2010, ApJ, 715, 1203\\
Murray, C.D. \& Dermott, S.F. 1999, \textit{Solar System Dynamics}, Cambridge UP, Cambridge\\
Patience, J. \etal 2002, ApJ, 581, 654\\
Raymond, S.N. \etal 2009, ApJ, 696, L98\\
------------. 2010, ApJ, 711, 772\\
Rivera, E.J. \& Lissauer, J.J. 2000, ApJ, 530, 454\\
Stepinski, T.F., Malhotra, R. \& Black, D.C. 2000, 545, 1044\\
Takeda, G. \& Rasio, F.A. 2005, ApJ, 627, 1001\\
Veras, D., \& Armitage, P. 2004, Icarus, 172, 349\\
Wolf, S. \& Klahr, H, 2002, ApJ, 578, L79\\
Wolszczan, A. 1994, Science, 264, 538\\
Wu, Y., \& Goldreich, P. 2002, ApJ, 564, 1024\\
Zhou, J.-L., \& Sun, Y.-S. 2003, ApJ, 598, 1290\\

\end{document}